# The cosmological constant: 2nd order quantum-mechanical correction to the Newton gravity

**Piero Chiarelli** [1,2,*]

[1] National Council of Research of Italy, Moruzzi 1, Pisa, 56124, Italy

[2] Interdepartmental Center "E. Piaggio" University of Pisa, Diotisalvi, 2, Pisa, 56122, Italy; pchiare@ifc.cnr.it; Phone: +39-050-315-2359; Fax: +39-050-315-2166

**Abstract:** The work shows that the associated Einstein-like gravity for the Klein-Gordon field shows the spontaneous emergence of the "cosmological" pressure tensor density (CPTD) that in the classical limit leads to the cosmological constant (CC). Even if the classical cosmological constant is set to zero, the model shows that exists a residual theory-derived quantum CPTD. The work shows that the cosmological constant can be considered as a second order quantum-mechanical correction to the Newtonian gravity. The outputs of the theory show that the expectation value of the CPTD is independent by the zero-point vacuum energy density and that it takes contribution only from the space where the mass is localized (and the space-time is curvilinear) while tending to zero as the space-time approaches to the flat vacuum. A developed model of scalar matter universe shows an overall cosmological effect of the CPTD on the motion of the galaxies that agrees with the astronomical observations.

**Keywords:** non-Minkowskian hydrodynamic representation of quantum equations; Einstein gravity of classical fields; energy-impulse tensor of classical Klein–Gordon field; cosmological constant

---

## 1. Introduction

Today, one of the most challenging problems in physics is the introduction of the space-time quantum mechanical properties into gravity.

Even if unable to give answers in a fully quantum regime, the semiclassical approximation [1–6] has brought successful results such as the explanation of the Hawking radiation and black hole (BH) evaporation [7].

The difficulties about the integration of QFT and the GE become really evident in the so called cosmological constant problem, a term that Einstein added to its equation to prevent the final collapse of the universe. The introduction, by hand, of the cosmological constant was then refused by Einstein himself who defined it as *the biggest mistake of my life* [8].

Actually, the CC has been reintroduced [1] to explain the astronomical observations about the motion of galaxies [9] and to give stability to the quantum vacuum, being the physical vacuum related to a positive, not null, CC [10-11].

Moreover, the energy-impulse tensor density (EITD) for classical bodies in the GE owns a point-dependence by the mass density (i.e., $\propto |Œ|^2$), without any analytically complete connection with the phase of the field $Œ = |Œ| \, exp \frac{i}{\hbar} S$. Due to this undefined connection between the GE and the particle fields, the integration between the QFT and the GE is still an open question that is the object of intense theoretical investigation.

Generally speaking, the link between the matter fields and the GE (as it happens for the photon field) can be obtained by defining an adequate GE. To this end, modifications to the Einstein equation have been proposed such as the scalar-tensor gravity and the Brans-Dicke modified gravity [12] that are able to solve the problem of the great disagreement between the measured and the theoretical values of the cosmological constant.

At glance with this point, the author has shown [13] that it is possible to obtain the GE, with analytical connection with the KGE field $\psi = |\psi| \, exp \frac{i}{\hbar} S$ , by using the quantum hydrodynamic description of a mass density $|\psi|^2$ owing the hydrodynamic impulse $\partial_\mu S = -p_\mu$ subject to the non-local quantum potential interaction [14-17] .

## 2. The Gravity Equation for the KGE on macroscopic scale

In reference [13] the author has shown that the gravity (with null classic CC) generated by the KGE field in the large-scale macroscopic limit (undergoing quantum decoherence) reads

$$R_{\mu\nu} - \frac{1}{2} R g_{\mu\nu} + \frac{8\pi G}{c^4} \Lambda g_{\mu\nu} = \frac{8\pi G}{c^4} \left( T_{(k)\mu\nu class} \left( 1 - \alpha_{(V_{qu(k)})} \right) - \Lambda_Q g_{\mu\nu} + \tau_{s\mu\nu(k)} \right) \qquad (1)$$

where [13]

$$T_{(k)\mu\nu class} = -|\psi_k|^2 c^2 \left( \frac{\hbar}{2} \right)^2 \left( (1-\alpha_{(V_{qu(k)})}) \frac{\frac{\hbar}{2i} \partial \, ln[\frac{\psi_k}{\psi_k *}]}{\partial t} \right)^{-1} \left( \frac{\partial \, ln[\frac{\psi_k}{\psi_k *}]}{\partial q^\mu} \frac{\partial \, ln[\frac{\psi_k}{\psi_k *}]}{\partial q^\nu} \right) \qquad (2)$$

$$V_{qu} = -\frac{\hbar^2}{m} \frac{1}{|\psi| / \sqrt{-g}} \partial_\mu \sqrt{-g} \left( g^{\mu\nu} \partial_\nu |\psi| \right) \qquad (3)$$

$$\Lambda_{(k)Q} = -|\psi_k|^2 \alpha_{(V_{qu1(k)})} \Delta_{||(k)} g_{\alpha\beta} \frac{mc^2}{\chi} u^\alpha u^\beta$$

$$= -|\psi_k|^2 \alpha_{(V_{qu1(k)})} \Delta_{||(k)} g_{\alpha\beta} c^2 \frac{i\hbar}{8} \frac{\frac{\partial \, ln[\frac{\psi_k}{\psi_k *}]}{\partial q_\alpha} \frac{\partial \, ln[\frac{\psi_k}{\psi_k *}]}{\partial q_\beta}}{\sqrt{1 - \frac{V_{qu\,(k)}}{mc^2}} \frac{\partial \, ln[\frac{\psi_k}{\psi_k *}]}{\partial t}} \qquad (4)$$

$$\tau_{s\mu\nu(k)} = -|\psi_k|^2 \alpha_{(V_{qu1(k)})} \Delta_{s\mu\nu(k)} g_{\alpha\beta} c^2 \frac{i\hbar}{2} \frac{\frac{\partial \, ln[\frac{\psi_k}{\psi_k *}]}{\partial q_\alpha} \frac{\partial \, ln[\frac{\psi_k}{\psi_k *}]}{\partial q_\beta}}{\sqrt{1 - \frac{V_{qu\,(k)}}{mc^2}} \frac{\partial \, ln[\frac{\psi_k}{\psi_k *}]}{\partial t}} \qquad (5)$$

$$\alpha_{(V_{qu(k)})} = \left( 1 - \sqrt{1 - \frac{V_{qu(k)}}{mc^2}} \right) \qquad (6)$$

and where $\Delta_{\}\}}$ and $\Delta_{s\mu\nu}$ are the isotropic and stress part, respectively, of the tensor

$\Delta_{\mu\nu} = \frac{\Delta_{\}\}}}{4} g_{\mu\nu} + \Delta_{s\mu\nu}$ that reads

$$\Delta_{\sim\text{€}(k)} = \frac{2}{r_{(V_{qu(k)})}} \left( \left( \frac{\partial}{\partial g^{\sim\text{€}}} - \frac{\partial}{\partial q^{\}}} \frac{\partial}{\frac{\partial g^{\sim\text{€}}}{\partial q^{\}}}} - \frac{\partial}{\partial /\text{Œ}_k/} \frac{\partial}{\frac{\partial g^{\sim\text{€}}}{\partial /\text{Œ}_k/}} \right) r_{(V_{qu(k)})} - \begin{pmatrix} \frac{\partial r_{(V_{qu(k)})}}{\partial q^{\}}} \frac{\partial ln\left(L_{(k)class}\right)}{\frac{\partial g^{\sim\text{€}}}{\partial q^{\}}}} + \frac{\partial ln\left(L_{(k)class}\right)}{\partial q^{\}}} \frac{\partial r_{(V_{qu(k)})}}{\frac{\partial g^{\sim\text{€}}}{\partial q^{\}}}} \\ \frac{\partial r_{(V_{qu(k)})}}{\partial /\text{Œ}_k/} \frac{\partial ln\left(L_{(k)class}\right)}{\frac{\partial g^{\sim\text{€}}}{\partial /\text{Œ}_k/}} + \frac{\partial ln\left(L_{(k)class}\right)}{\partial /\text{Œ}_k/} \frac{\partial r_{(V_{qu(k)})}}{\frac{\partial g^{\sim\text{€}}}{\partial /\text{Œ}_k/}} \end{pmatrix} \right) \tag{7}$$

where

$$L_{(k)class} = c^2 \frac{i\hbar}{2} \left( r_{(V_{qu(k)})} \frac{\partial ln[\frac{\text{Œ}_k}{\text{Œ}_k *}]}{\partial t} \right)^{-1} g_{\sim r} \frac{\partial ln[\frac{\text{Œ}_k}{\text{Œ}_k *}]}{\partial q_r} \frac{\partial ln[\frac{\text{Œ}_k}{\text{Œ}_k *}]}{\partial q_{\text{€}}}. \tag{8}$$

### 2.1. Perturbative approach to the GE–KGE evolutionary system of equations

For particles very far from the Planckian mass density $\frac{m_p}{l_p^{3}} = \frac{c^5}{\hbar G^2}$, it is possible to solve the system of equations

$$R_{\sim\text{€}} - \frac{1}{2} R g_{\sim\text{€}} + \frac{8f\,G}{c^4} \Lambda g_{\sim\text{€}} = \frac{8f\,G}{c^4} \begin{pmatrix} T_{(k)\sim\text{€}class} \left(1 - r_{(V_{qu(k)})}\right) \\ -r_{(V_{qu(k)})} \begin{pmatrix} \left( /\text{Œ}_k/^2 \, L_{(k)class} \frac{\Delta_{\}\}}}{4} \right) g_{\sim\text{€}} \\ + /\text{Œ}_k/^2 \, L_{(k)class} \Delta_{s\sim\text{€}(k)} \end{pmatrix} \end{pmatrix} \tag{9}$$

$$\left( g^{\sim\text{€}} \partial_{\text{€}} \text{Œ} \right)_{;\sim} = -\frac{m^2 c^2}{\hbar^2} \text{Œ} \tag{10}$$

by a perturbative iteration,

$$R_{\sim\text{€}\left(v_{\sim\text{€}}\right)} \cong R^{(0)}{}_{\sim\text{€}\left(v_{\sim\text{€}}\right)} + R^{(1)}{}_{\sim\text{€}\left(v_{\sim\text{€}}\right)} + R^{(2)}{}_{\sim\text{€}\left(v_{\sim\text{€}}\right)} + .... \tag{11a}$$

$$\text{Œ} = \text{Œ}_0 + \text{Œ}' + \text{Œ}'' + ..... \tag{11b}$$

$$g_{\text{€}\sim} = y_{\text{€}\sim} + v_{\sim\text{€}} = \begin{bmatrix} 1 & 0 & 0 & 0 \\ 0 & -1 & 0 & 0 \\ 0 & 0 & -1 & 0 \\ 0 & 0 & 0 & -1 \end{bmatrix} + h^{(1)}_{\sim\text{€}} + h^{(2)}_{\sim\text{€}} + ... \tag{12}$$

where$(V_{\sim\text{€}}V^{\sim\text{€}} = |V|^2 << 1)$ and where $y_{\text{€}\sim}$ satisfies the static solution $R^{(0)}_{\sim\text{€}\,(y_{\text{€}\sim})} = 0, R^{(0)}_{(y_{\text{€}\sim})} = 0$, of the zero order GE

$$R^{(0)}_{\sim\text{€}} - \frac{1}{2} R^{(0)} = 0. \tag{13}$$

$\text{Œ}_0$ is the solution of the zero order KGE,

$$\partial_{\sim}\partial^{\sim}\text{Œ}_0 = -\frac{m^2c^2}{\hbar^2}\text{Œ}_0 \tag{14}$$

$\text{Œ}'$ the solution of the first order KGE,

$$\partial_\sim\partial^\sim \text{Œ}' + \frac{m^2c^2}{\hbar^2}\text{Œ}' = -\partial_\sim h^{(1)\sim\text{€}}\partial_{\text{€}}\text{Œ} \cong -\partial_\sim h^{(1)\sim\text{€}}\partial_{\text{€}}\text{Œ}_0 \tag{15}$$

(at first order $V_{\sim\text{€}} = h^{(1)}_{\sim\text{€}}$) and $h^{(1)}_{\sim\text{€}}$ is the solution of the first order GE,

$$R^{(1)}_{\sim\text{€}\left(h^{(1)}{}_{\sim\text{€}}\right)} - \frac{1}{2}R^{(1)}_{\sim\text{€}\left(h^{(1)}{}_{\sim\text{€}}\right)} g_{\sim\text{€}} = \frac{8f G}{c^4}\begin{pmatrix} T^{(0)}{}_{(k)\sim\text{€}\,class}\left(1-\text{r}^{(0)}{}_{(V_{qu(k)})}\right) \\ -\text{r}^{(0)}{}_{(V_{qu(k)})}\begin{pmatrix}\left(|\text{Œ}_{0k}|^2 L^{(0)}{}_{(k)class}\frac{\Delta_{\}\}}}{4}\right) g_{\sim\text{€}} \\ +|\text{Œ}_{0k}|^2 L^{(0)}{}_{(k)class}\Delta_{s\sim\text{€}(k)}\end{pmatrix}\end{pmatrix} \tag{16}$$

where the Christoffel symbol reads [18]:

$$\Gamma^{\text{r}}_{\text{€}\sim} = \frac{1}{2} y^{\text{rs}}\left(\partial_\sim V_{s\text{€}} + \partial_{\text{€}} V_{s\sim} - \partial_s V_{\text{€}\sim}\right) \tag{17}$$

leading to

$$R_{\sim\text{€}(h_{\sim\text{€}})} = \left(\partial_l\Gamma^l_{\sim\text{€}} - \partial_{\text{€}}\Gamma^l_{\sim l} + \Gamma^l_{\sim\text{€}}\Gamma^m_{lm} - \Gamma^m_{\sim l}\Gamma^l_{\text{€}m}\right) \tag{18}$$

Moreover, by using the zero-order relations

$$V_{qu_0(k)} = 0 \tag{19}$$

and

$$\text{r}^{(0)}{}_{(V_{qu_0(k)})} = 0 \tag{20}$$

it follows that (16) the first-order GE reads

$$R^{(1)}_{\sim\text{€}\left(h^{(1)}{}_{\sim\text{€}}\right)} - \frac{1}{2}R^{(1)}_{\sim\text{€}\left(h^{(1)}{}_{\sim\text{€}}\right)} g_{\sim\text{€}} = \frac{8f G}{c^4} T^{(0)}{}_{(k)\sim\text{€}\,class} = \frac{8f G}{c^4}|\text{Œ}_{\tilde{k}}|^2 \frac{mc^2}{\text{x}} u^{(0)}{}_\sim u^{(0)}{}_{\text{€}} \tag{21}$$

from which we can readily see that the GE weak gravity limit, on macroscopic scale at the first-order, coincides with the general relativity equation leading to the Newtonian potential of gravity. The first contribution to the cosmological constant comes from the second order of approximation

$$R^{(2)}_{\sim\text{€}\left(V_{\sim\text{€}}\right)} - \frac{1}{2}R^{(2)}_{\left(V_{\sim\text{€}}\right)} g_{\sim\text{€}} = \frac{8f G}{c^4}\left(\left(1-\text{r}^{(1)}_{(\tilde{k})}\right)T^{(1)}{}_{(k)\sim\text{€}\,class} - \Lambda^{(1)}_{(\tilde{k})} g_{\sim\text{€}} + \Delta\ddagger^{(1)dec}_{\sim\text{€}\,stress}\right) \tag{22}$$

where

$$T^{(1)}{}_{(k)\sim\text{€}\,class} = /\text{Œ}_k/^2 \frac{mc^2}{\mathrm{x}} u^{(1)}{}_{\sim} u^{(1)}{}_{\text{€}} \tag{23}$$

$$\Lambda^{(1)}_{(k)Q} = -/\text{Œ}_k/^2 \,\mathrm{r}^{(1)}_{(V_{qu_1(k)})} \Delta^{(1)}_{||(k)} g_{\mathrm{rs}} c^2 \frac{i\hbar}{8} \frac{\frac{\partial\, ln[\frac{\text{Œ}_k}{\text{Œ}_k{}^*}]}{\partial q_{\mathrm{r}}} \frac{\partial\, ln[\frac{\text{Œ}_k}{\text{Œ}_k{}^*}]}{\partial q_{\mathrm{s}}}}{\sqrt{1-\frac{V_{qu\ (k)}}{mc^2}}\ \frac{\partial\, ln[\frac{\text{Œ}_k}{\text{Œ}_k{}^*}]}{\partial t}}$$

$$\cong -/\text{Œ}_k/^2 \,\mathrm{r}^{(1)}_{(V_{qu_1(k)})} \frac{\Delta^{(1)}_{||(k)}}{4} g_{\mathrm{rs}} \frac{mc^2}{\mathrm{x}} u^{\mathrm{r}}_{class} u^{\mathrm{s}}_{class} \tag{24}$$

$$\cong -/\text{Œ}_k/^2 \,\mathrm{r}^{(1)}_{(V_{qu_1(k)})} \frac{\Delta^{(1)}_{||(k)}}{4} \mathrm{y}_{\mathrm{rs}} \frac{mc^2}{\mathrm{x}} u^{\mathrm{r}}_{class} u^{\mathrm{s}}_{class}$$

$$\cong -/\text{Œ}_k/^2 \,\mathrm{r}^{(1)}_{(V_{qu_1(k)})} \frac{\Delta^{(1)}_{||(k)}}{4} \frac{mc^2}{\mathrm{x}}$$

where it has been used the property (near the Minkowskian limit)

$$\mathrm{y}_{\mathrm{rs}} u^{\mathrm{r}}_{class} u^{\mathrm{s}}_{class} = 1 \tag{25}$$

$$\Delta\ddagger^{(1)dec}_{\sim\text{€}\,stress} = -/\text{Œ}_k/^2 \,\mathrm{r}^{(1)}_{(V_{qu_1(k)})} \Delta^{(1)}_{s\sim\text{€}(k)} g_{\mathrm{rs}} c^2 \frac{i\hbar}{2} \frac{\frac{\partial\, ln[\frac{\text{Œ}_k}{\text{Œ}_k{}^*}]}{\partial q_{\mathrm{r}}} \frac{\partial\, ln[\frac{\text{Œ}_k}{\text{Œ}_k{}^*}]}{\partial q_{\mathrm{s}}}}{\sqrt{1-\frac{V_{qu\ (k)}}{mc^2}}\ \frac{\partial\, ln[\frac{\text{Œ}_k}{\text{Œ}_k{}^*}]}{\partial t}}$$

$$\cong -/\text{Œ}_k/^2 \,\mathrm{r}^{(1)}_{(V_{qu_1(k)})} \Delta^{(1)}_{s\sim\text{€}(k)} g_{\mathrm{rs}} \frac{mc^2}{\mathrm{x}} u^{\mathrm{r}}_{class} u^{\mathrm{s}}_{class} \tag{26}$$

$$\cong -/\text{Œ}_k/^2 \,\mathrm{r}^{(1)}_{(V_{qu_1(k)})} \Delta^{(1)}_{s\sim\text{€}(k)} \mathrm{y}_{\mathrm{rs}} \frac{mc^2}{\mathrm{x}} u^{\mathrm{r}}_{class} u^{\mathrm{s}}_{class}$$

$$\cong -/\text{Œ}_k/^2 \,\mathrm{r}^{(1)}_{(V_{qu_1(k)})} \Delta^{(1)}_{s\sim\text{€}(k)} \frac{mc^2}{\mathrm{x}}$$

Moreover, by introducing in (23) the approximation

$$\text{Œ} \cong \text{Œ}_0 + \text{Œ}\,' \tag{27}$$

it follows that

$$\frac{mc^2}{\mathrm{x}} u^{(1)\mathrm{r}} u^{(1)\text{€}} = c^2 \frac{i\hbar}{2} \frac{\frac{\partial\, ln[\frac{\text{Œ}_k}{\text{Œ}_k{}^*}]}{\partial q_{\mathrm{r}}} \frac{\partial\, ln[\frac{\text{Œ}_k}{\text{Œ}_k{}^*}]}{\partial q_{\text{€}}}}{\sqrt{1-\frac{V_{qu(k)}}{mc^2}}\ \frac{\partial\, ln[\frac{\text{Œ}_k}{\text{Œ}_k{}^*}]}{\partial t}} = \frac{mc^2}{\mathrm{x}} \left( u^{\mathrm{r}(0)} u^{\text{€}(0)} + \Delta u^{\mathrm{r}} \Delta u^{\text{€}} \right) \tag{28}$$

where,

$$\frac{mc^2}{x} u^{r(0)} u^{€(0)} = c^2 \frac{i\hbar}{2} \frac{ \frac{\partial ln[\frac{Œ_{0k}}{Œ_{0k}*}]}{\partial q_r} \; \frac{\partial ln[\frac{Œ_{0k}}{Œ_{0k}*}]}{\partial q_€} }{ \sqrt{1-\frac{V_{qu0(k)}}{mc^2}} \; \frac{\partial ln[\frac{Œ_{0k}}{Œ_{0k}*}]}{\partial t} } = c^2 \frac{k^r k^€}{Š_k} \tag{29}$$

where, at first order,

$$\begin{aligned} \frac{mc^2}{x} \Delta u^r \Delta u^€ \cong\; & c^2 \frac{k^r k^€}{Š_k{}^2} \left( -\partial_t \left( \frac{Œ'_k}{Œ_{0k}} - \frac{Œ'_k{}^*}{Œ_{0k}{}^*} \right) + \frac{Š_k V^{(1)}_{qu(Œ'_k)}}{2mc^2} \right) \\ & + c^2 \frac{i\hbar}{2Š_k} \left( k^r \partial^€ \left( \frac{Œ'_k}{Œ_{0k}} - \frac{Œ'_k{}^*}{Œ_{0k}{}^*} \right) + \partial^r \left( \frac{Œ'_k}{Œ_{0k}} - \frac{Œ'_k{}^*}{Œ_{0k}{}^*} \right) k^€ \right) \end{aligned} \tag{30}$$

If the macroscopic quantum-mechanical GE (1) and the Einstein equation coincide themselves at first order, the second order of the GE (1) contains contributions to the gravity (among those the cosmological isotropic pressure $\Lambda^{tot}_{(k)Q}$) that go to zero if $\hbar$ is set to zero showing to be quantum-mechanical correction to the Newtonian weak gravity.

It is worth mentioning that the GE (1) shows the additional correction to the Newton gravity $\Delta ‡^{(1)dec}_{~€ stress}$ that, being the product of two second order terms, $r^{(1)}_{(V_{qu1(k)})}$ and $\Delta^{(1)}_{s~€(k)}$, that increase at high-curvature space-time (i.e., $V_{qu} \approx mc^2$ , and $r_{(V_{qu(k)})} \approx max\{ r_{(V_{qu(k)})} \} = 1$ in black holes (see section 4.1) ), should give well detectable effects near black holes such as, for instance, in the motion of stars in the bulge around the supermassive BH (SMBH) at the center of the galaxies or in the rotation of twin black holes and neutron stars.

## 3. The cosmological pressure tensor density expectation-value of the quantum KGE field in the quasi-Minkowskian space-time

At zero order, very far from the Planckian mass densities, the GE (1) leads to a Minkowskian KGE field and, hence, when the KGE field is quantized, to the standard QFT outputs.

The quantization of the GE–KGE system of equations is not the objective of this work, but it is interesting to evaluate the CPTD expectation value $< 0_k / \Lambda^{tot}_{(k)Q} g_{~€} / 0_k >$ and to see if it can give a compatible output with the observed CC .

To this end, we need to express the quantum potential as a function of the annihilation and creation operators $a_{(k)}$ and $a^{\dagger}{}_{(k)}$ of the free KGE quantum field

$$Œ = \iint \frac{d^3k}{(2ƒ)^3} \frac{1}{2Š_k} \left( a_{(k)} exp[ik_r q^r] + a^{\dagger}{}_{(k)} exp[-ik_r q^r] \right) \tag{31}$$

that, by using the discrete form of field Fourier decomposition (i.e., $\iint \frac{d^3k}{(2ƒ)^3} \to \frac{1}{\sqrt{V}} \sum_{k=0}$ ) reads

$$Œ = \frac{1}{\sqrt{V}} \sum_{k=0}^{\infty} \frac{1}{2Š_k} \left( a_{(k)} exp[\frac{ip_r q^r}{\hbar}] + a^{\dagger}{}_{(k)} exp[-\frac{ip_r q^r}{\hbar}] \right) \tag{32}$$

where $\frac{p_{\sim}}{\hbar} = k_{\sim}$, leads to

$$
\begin{aligned}
V_{qu0(k>0)} &= -\frac{\hbar^2}{m}\frac{1}{/\text{Œ}_k/\sqrt{-g}}\partial_{\sim}\sqrt{-g}\,g^{\sim\text{€}}\partial_{\text{€}}\,/a(k)exp[ik_{\mathrm{r}}q^{\mathrm{r}}]/ \\
&= -\frac{\hbar^2}{m}\frac{1}{/\text{Œ}_k/}\partial_{\sim}\partial^{\sim}\,/a(k)exp[ik_{\mathrm{r}}q^{\mathrm{r}}]/ \qquad (33)\\
&= -\frac{\hbar^2}{m}\frac{1}{/\text{Œ}_k/}\partial_{\sim}\partial^{\sim}\sqrt{a(k)a^{\dagger}(k)} = V_{qu0(k<0)}
\end{aligned}
$$

where, now, $a(k), a^{\dagger}(k)$ obey to the commutation relations $[a(k), a^{\dagger}(k')] = \mathrm{u}_{kk'}$

However, even if the squared root of operators $\sqrt{a(k), a^{\dagger}(k)}$ in the quantum potential (33) can be defined by making use of the Taylor expansion series, the higher order terms own the problem to be ordered. In order to remark this freedom in the definition of the quantum potential operator, we name it as $\hat{V}_{qu}^{Q-ord}$ and: reads

$$
\hat{V}_{qu0}^{Q-ord} = -\frac{\hbar^2}{m}\frac{1}{\left(\sqrt{a(k)a^{\dagger}(k)}\right)^{Q-ord}\sqrt{-g}}\partial_{\sim}\sqrt{-g}\,g^{\sim\text{€}}\partial_{\text{€}}\left(\sqrt{a(k)a^{\dagger}(k)}\right)^{Q-ord} \qquad (34)
$$

Moreover, by using the identity [13]

$$
\frac{\partial S}{\partial q^{\sim}}\frac{\partial S}{\partial q_{\sim}} = p_{\sim}p^{\sim} = \left(\frac{E^2}{c^2} - p^2\right) = m^2c^2\left(1 - \frac{V_{qu}}{mc^2}\right) \qquad (35)
$$

and the identities $p_{\sim} = -\partial_{\sim}S = (p_0, -p_i)$ and $p^2 = p_i p_i$, it follows that:

$$
\check{S}_k{}^2 = \frac{c^2p^2}{\hbar^2} + \frac{m^2c^4}{\hbar^2}\left(1 - \frac{\hat{V}_{qu}^{Q-ord}}{mc^2}\right) \qquad (36)
$$

and the CPTD expectation value reads

$$
<0_k/\Lambda_{(k)}^{tot}/0_k> = \frac{1}{(2f)^2\hbar c}\int^{\check{S}_{k_{max}}}\Lambda_{(k)}^{tot}\sqrt{\frac{\hbar^2\check{S}_k{}^2}{c^2} - m^2c^2\left(1 - \frac{\hat{V}_{qu}^{Q-ord}}{mc^2}\right)}\,d\check{S}_k \qquad (37)
$$

where $/0_k>$ represents the k-indexed harmonic oscillators of the field in the fundamental state [19].

Moreover, being $a(k)$ just a function of the moment, in the Minkowskian limit, we obtain $\partial^{\sim}\left(\sqrt{a(k)a^{\dagger}(k)}\right)^{Q-ord} = 0$, and hence that

$$
\hat{V}_{qu0}^{Q-ord} = 0 \qquad (38)
$$

$$
\mathrm{r}_{(\hat{V}_{qu0}^{Q-ord})} = 0 \qquad (39)
$$

$$
\Delta_{||(k)} = 0 \qquad (40)
$$

$$
\Lambda_{(k)}^{tot} = 0 \qquad (41)
$$

$$
<0_p/\Lambda_{(k)}^{tot}/0_p> = -\frac{1}{(2f)^2\hbar c}\int^{\check{S}_{k_{max}}}\left(\mathrm{r}^{(1)}_{(\hat{V}_{qu0}^{Q-ord})}\frac{\Delta_{||(k)}^{(1)}}{4}\frac{mc^2}{\mathrm{x}}\right)\sqrt{\frac{\hbar^2\check{S}_k{}^2}{c^2} - m^2c^2}\,d\check{S}_k = 0 \qquad (42)
$$

and, finally, from (38), that

$$\omega_k{}^2 = \frac{c^2 p^2}{\hbar^2} + \frac{m^2 c^4}{\hbar^2} = c^2 k^2 + \frac{m^2 c^4}{\hbar^2} . \qquad (43)$$

From the last identity, the standard QFT outputs are warranted at the zero order of gravitational approximation.

## 4. The cosmological constant of the quantum-mechanical GE

If in the classical Einstein equation the "cosmological" constant $\Lambda$ (a factor multiplying the metric tensor $g_{\mu\nu}$) cannot be function of the space [20], on the contrary, in the GE (1) the CPTD $\Lambda g_{\mu\nu}$ is both a function of the mass distribution density $|\psi|^2$ and of the quantum potential $V_{qu}$.

As shown by in the section 3, the CC is different from zero only in the regions of curved space-time where the mass is localized (e.g., $\frac{\partial |\psi|}{\partial q^{\nu}} \neq 0$) and it higher larger is the gravitational field (while it is null in the vacuum very far from massive objects given that if $\frac{\partial |\psi|}{\partial q^{\nu}} = 0$ also $V_{qu} = 0$).

Thence, the effect on cosmological motion of galaxies the term $\Lambda g_{\mu\nu}$ given by the mean value $< \Lambda >$ over all the volume of the galactic space $V_g$, that reads

$$< \frac{\Lambda}{c^2} > = \frac{1}{V_g} \iiint_{V_g} \frac{\Lambda^{tot}_{(k)}}{c^2} dV , \qquad (44)$$

takes contributions just from the places where matter is present (e.g., particles core and black holes (BH)).

### 4.1. The toy model: the cosmological constant of the universe whose mass is distributed in scalar bosons of Plank's mass

In order to fully evaluate expression (44), we should introduce the electromagnetic, the strong, the weak forces and the Higgs boson interaction that give rise to the family of elemental particles.

For the gravity of classical fields developed so far (the classical KGE field coupled to the gravitational force) the localization of the mass can only come from gravity whose stable localized states are those of BH with a mass larger than the Planck mass [21].

In order to evaluate (44) by using this oversimplified model, we represent all the mass of the universe (baryonic matter, dark matter and dark energy, respectively) by BHs particles of Plancknian mass $m$ (i.e., Plancknian Black hole (PBH)) with the volume of a sphere of Planck radius $\Delta V_{PBH} \approx 2 \; 10^{-105} m^3$. Thence, by posing that

$$< \frac{\Lambda}{c^2} > \approx \frac{1}{V_g} \sum_{i=1}^{N} \begin{pmatrix} < \frac{\Lambda}{c^2} >_{barionic} + < \frac{\Lambda}{c^2} >_{dark\ m} \\ + < \frac{\Lambda}{c^2} >_{dark\ En} \end{pmatrix} \Delta V_{i_{PBH}} = \frac{\Delta V_{PBH}}{V_g} \sum_{i} \begin{pmatrix} < \frac{\Lambda_i}{c^2} >_{barionic} + < \frac{\Lambda_i}{c^2} >_{dark\ m} \\ + < \frac{\Lambda_i}{c^2} >_{dark\ En} \end{pmatrix} \qquad (45)$$

where $V_g$ is the universe volume, and by utilizing (4) it follows that

$$< \frac{\Lambda}{c^2} >_{barionic} + < \frac{\Lambda}{c^2} >_{dark\ m} \approx \frac{\mathrm{u}V_{_{PBH}}}{V_g} m \sum_i < \frac{\mathrm{r}_{(V_{qu1(k)})}\Delta_{|\,|(k)}}{4} \frac{/\text{Œ}\,/_i^2}{\mathrm{x}} \left( 1 - \sqrt{1 - \frac{V_{qu}}{mc^2}} \right) >$$

$$= \frac{\mathrm{u}V_{_{PBH}}}{V_g} m < \frac{\mathrm{r}_{(V_{qu1(k)})}\Delta_{|\,|(k)}}{4} \frac{/\text{Œ}\,/_{_{PBH}}^2}{\mathrm{x}} \left( 1 - \sqrt{1 - \frac{V_{qu}}{mc^2}} \right) \quad (46)$$

where $/\text{Œ}\,/_{_{PBH}}^2 = \frac{\sum_i /\text{Œ}\,/_i^2}{N}$ is the mean number of Planckian black holes per volume of the universe and $m = m_{barionic} + m_{dark\ m}$, where $m_{barionic}\mathrm{u}V_{pBH}$ and $m_{dark\ m}\mathrm{u}V_{pBH}$ are the baryonic and dark matter mass, respectively, contained in each PBH.

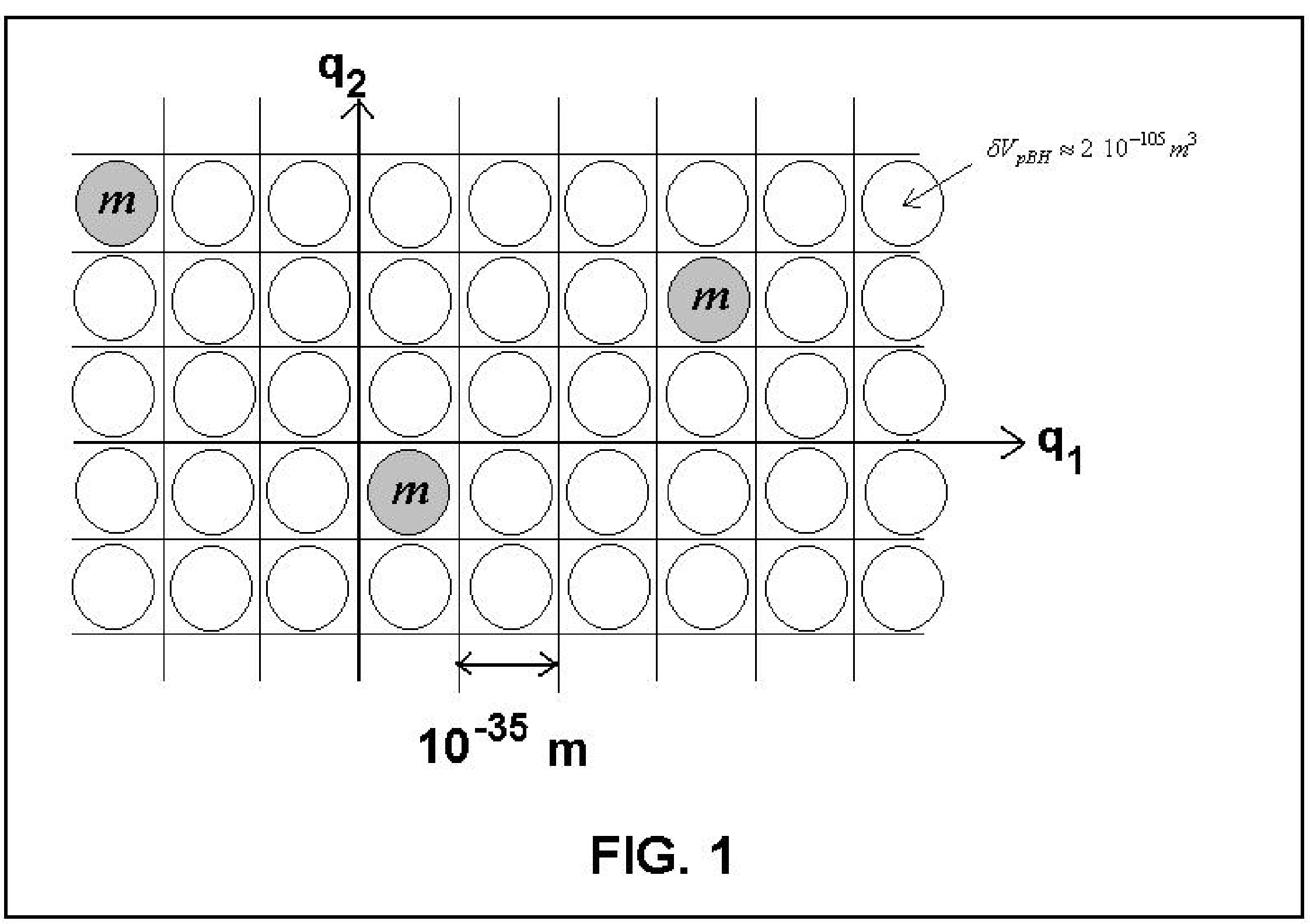

**Figure 1.** Figure 1 shows the distribution the Plancknian cells of vacuum of (baryonic and dark) mass $m$

Figure 1 shows the discretization used to evaluate (44). Here the vacuum is assumed composed of cells of volume $\mathrm{u}V_{pBH} \approx 2\ 10^{-105} m^3$ whose radius is of order of the Planck length. Some of them contain the localized mass $m$ of the PBH of baryonic and dark matter.

Moreover, assuming that he zero-point energy density of the vacuum $E_0$ represents the dark energy we add it, or the equivalent mass $m_{dark\ En} = \frac{E_0}{c^2}$, to each cell of the space, to obtain

$$< \frac{\Lambda_i}{c^2} >_{dark\ En} \approx \frac{\text{u} V_{PBH}}{V_g} \frac{E_0}{c^2} < \frac{\text{r}_{(V_{qu_1(k)})} \Delta_{||(k)}}{4} \frac{\text{/Œ /}^2_{PBH_{dark\ En}}}{\text{x}} \left( 1 - \sqrt{1 - \frac{V_{qu}}{mc^2}} \right) > \qquad . \tag{47}$$

leading to the representative set up in figure 2

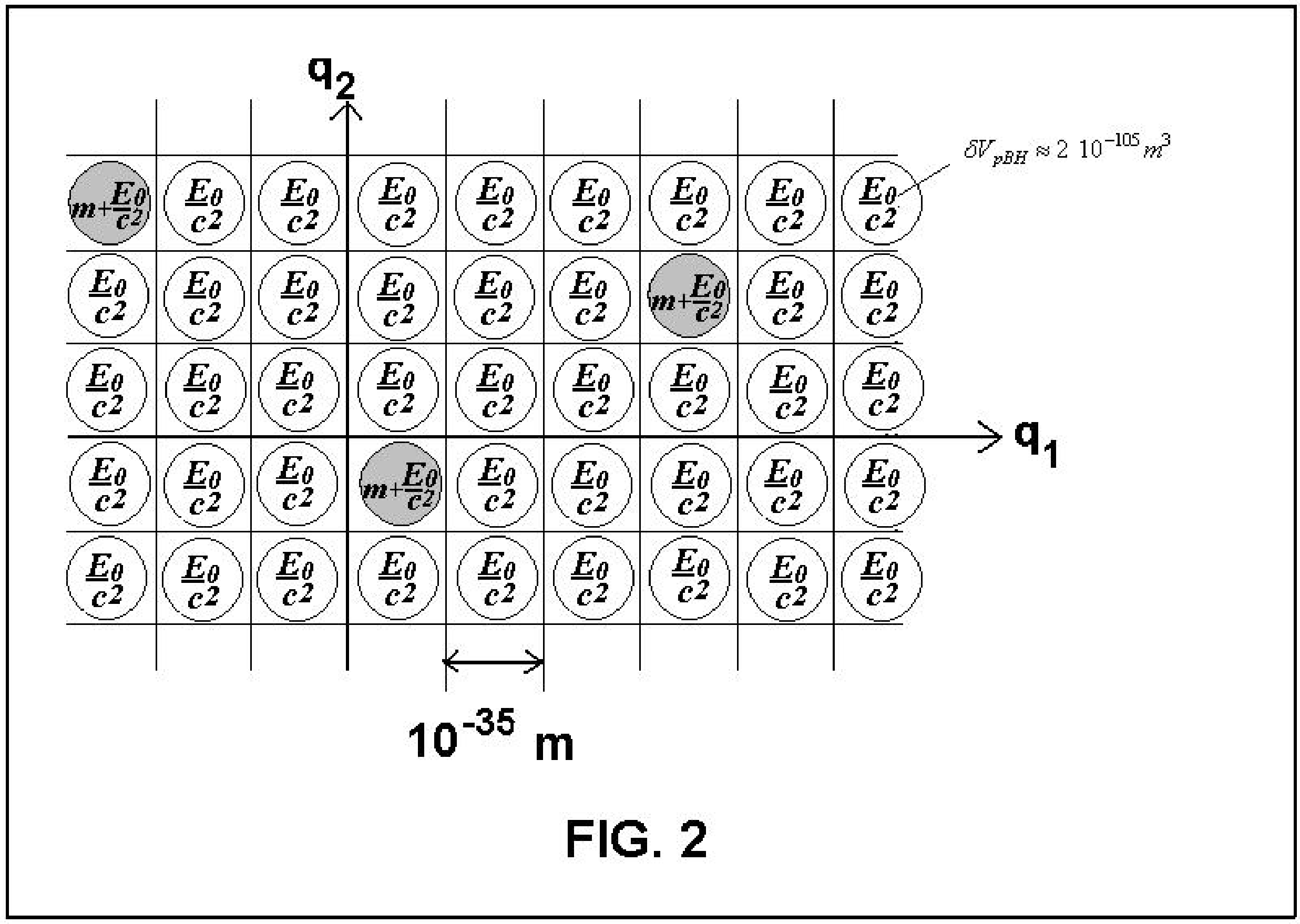

**Figure 2.** The figure 2 shows the baryonic mass, dark mass and dark energy distribution in the Plancknian cells of the vacuum

Moreover, given that the constant mass distribution in the vacuum $\text{/Œ /}_{PBH_{tot}} = \text{/Œ /}_{PBH_{barionic}} + \text{/Œ /}_{PBH_{dark\ m}} + \text{/Œ /}_{PBH_{dark\ En}}$ does not give contribution both to the quantum potential (33) and to the cosmological constant (4), it follows that in (44) only the places where there are the baryonic and the dark matter leads to a not null contribution to the cosmological constant.

Thence, the number of PBHs per volume $\text{/Œ /}^2_{PBH_{dark\ En}}$, that contain the dark energy contribution to the cosmological constant, are equal to the number of PBH per volume of the baryonic and dark matter $\text{/Œ /}^2_{PBH_m}$.

Thence, from (46-47) It follows that

$$<\frac{\Lambda}{c^2}>\approx\frac{\mathrm{u}V_{PBH}}{V_g}m_{tot}<\frac{\mathrm{r}_{(V_{qu_1(k)})}\Delta_{||(k)}}{4}\frac{\text{/Œ ʃ}^2_{PBH_m}}{\mathrm{x}}\left(1-\sqrt{1-\frac{V_{qu}}{mc^2}}\right)> \qquad (48)$$

where $m_{tot}=m+\frac{E_0}{c^2}$ .

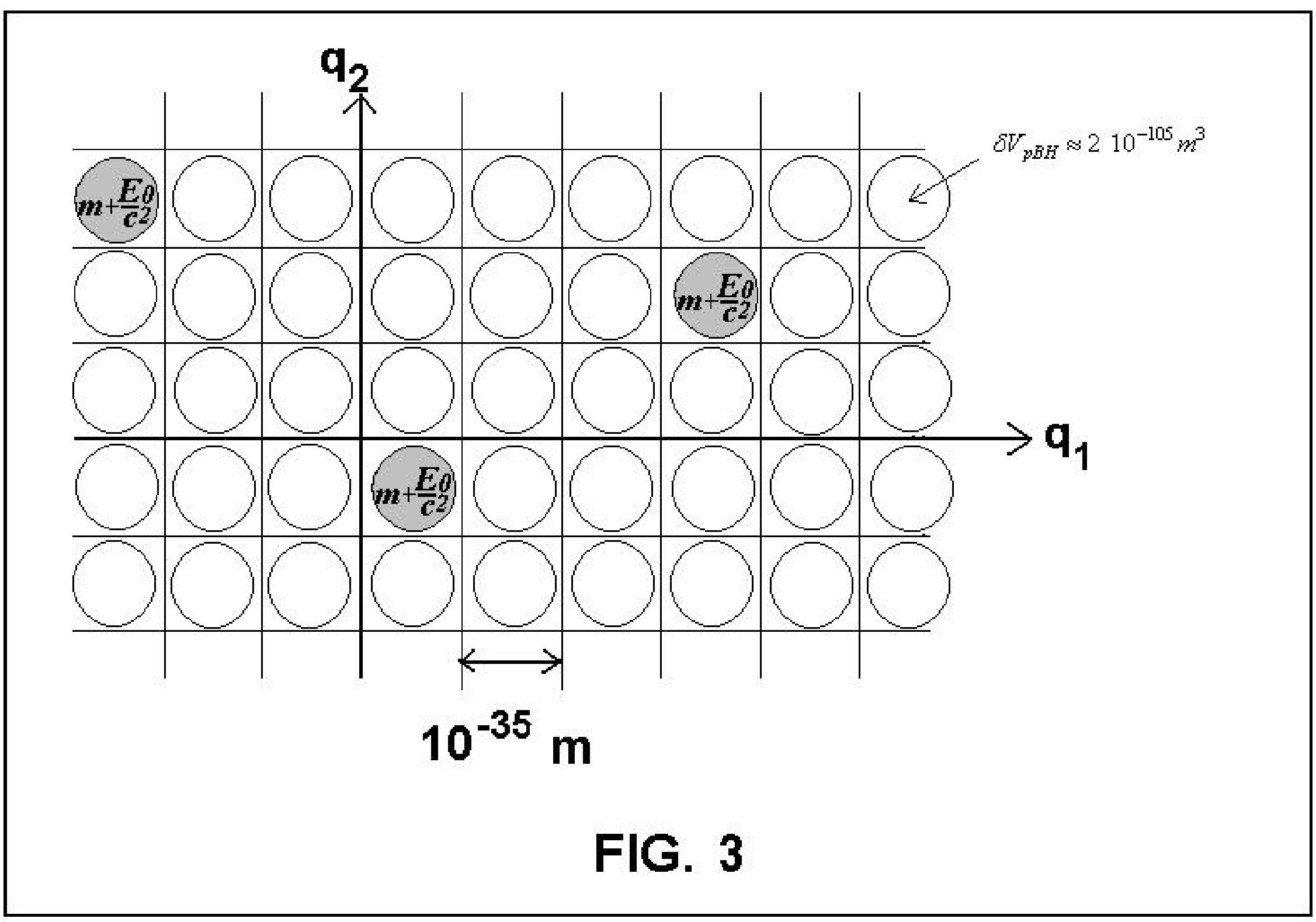

**Figure 3.** Figure3 shows the Plancknian cells of the vacuum contributing to the quantum potential and, therefore, to the cosmological pressure tensor density

Moreover, by posing both that

$$<\frac{\text{/Œ ʃ}^2_{PBH}}{\mathrm{x}}\left(1-\sqrt{1-\frac{V_{qu}}{mc^2}}\right)>\approx<\text{/Œ ʃ}^2_{PBH}><\frac{\mathrm{r}_{(V_{qu_1(k)})}\Delta_{||(k)}}{4}\frac{1}{\mathrm{x}}\left(1-\sqrt{1-\frac{V_{qu}}{mc^2}}\right)> \qquad (49)$$

that by using the relation

$$V_{mean}=\frac{V_g}{<\text{/Œ ʃ}^2_{PBH}>}, \qquad (50)$$

where $V_{mean}$ is the galactic space, available per PBH mass it follows that

$$< \frac{\Lambda}{c^2} > \approx \frac{\left( m_{b+d} + \frac{E_0}{c^2} \right) \mathrm{u} V_{pBH}}{V_{mean}} < \frac{\mathrm{r}_{(V_{qu1(k)})} \Delta_{||(k)}}{4} \frac{1}{\mathrm{x}} \left( 1 - \sqrt{1 - \frac{V_{qu}}{mc^2}} \right) >, \qquad (51)$$

where $V_{mean} \approx 2\ 10^5 m^3$ [22] (for the baryonic mass, 4% of the total mass, leads to about eight hydrogen atoms per cubic meter) .

Furthermore, in order to evaluate the order of magnitude of $< \frac{\Lambda}{c^2} >$, we introduces the experimental data both for the baryonic and dark matter density $\Omega_m$ and for the dark energy density $\Omega_{dark\ En}$ in the expression (51) to obtain

$$\begin{aligned} < \frac{\Lambda}{c^2} > &\approx \frac{\left( m_{b+d} + \frac{E_0}{c^2} \right) \mathrm{u} V_{pBH}}{V_{mean}} \frac{< \frac{\mathrm{r}_{(V_{qu1(k)})} \Delta_{||(k)}}{4} >}{< \mathrm{x} >} < \left( 1 - \sqrt{1 - \frac{V_{qu}}{mc^2}} \right) > \\ &\approx \left( \Omega_m + \Omega_{dark\ En} \right) \frac{< \frac{\mathrm{r}_{(V_{qu1(k)})} \Delta_{||(k)}}{4} >}{< \mathrm{x} >} \left( 1 - \sqrt{1 - \frac{< V_{qu} >}{mc^2}} \right) \end{aligned} \qquad (52)$$

where [22]

$$\Omega_m = \frac{m_{b+d} \mathrm{u} V_{pBH}}{V_{mean}} \cong 2{,}6\ 10^{-27} kg / m^3, \qquad (53)$$

and where

$$\Omega_{dark\ En} = \frac{\frac{E_0}{c^2} \mathrm{u} V_{pBH}}{V_{mean}} \mathrm{u} V_{pBH} \cong 7{,}4\ 10^{-27} kg / m^3 . \qquad (54)$$

For low velocity massive particles (e.g., moving at speed lower than $\frac{1}{3}c$ ) we can assume

$$< \mathrm{x} > \approx \frac{1}{\sqrt{1 - \frac{< \dot{q}^2 >}{c^2}}} \approx 1 . \qquad (55)$$

As far as it concerns $< V_{qu} >$, we evaluate it for a Schwarzschild radial symmetric BH.

Given the metric tensors in the spherical coordinates

$$\begin{aligned} &g_{00} = e^{€} ; g_{11} = -e^{\}} ; g_{22} = -r^2 ; \\ &g_{33} = -r^2 \sin^2 „ ; \ \sqrt{-g} = / e^{\frac{\}+€}{2}} r^4 \sin^2 „ /^{-1} \end{aligned}, \qquad (56)$$

for the Schwarzschild BH, they read [21, 23]

$$g_{11} = -e^{\}} = -g_{00}^{-1} \qquad (57)$$

$$g_{00} = e^{€} = e^{-\}} = -\left(1-\frac{R_g}{r}\right) \qquad (58)$$

$$g = -r^4 \, sin^2 \, [ \; . \qquad (59)$$

By introducing (57-59), in the hydrodynamic equation of motion [13] (for the radial co-ordinate) that by using the stationary conditions for eigenstates

$$\frac{du_{\tilde{}}}{dt} = 0 \qquad (60)$$

$$u_{\tilde{}} = (1,0,0,0) \, , \qquad (61)$$

reads

$$\frac{du_{\tilde{}}}{dt} = 0 = \frac{1}{2}\frac{c}{\mathrm{X}} u^{\}} u^{|} \frac{\partial g_{\}|}}{\partial q^{\tilde{}}} + \frac{c}{\mathrm{X}} \frac{\partial \, ln \sqrt{1-\frac{V_{qu(n)}}{mc^2}}}{\partial q^{\tilde{}}} + \frac{c}{\mathrm{X}} \frac{\partial \, ln \mathrm{X}}{\partial q^{\tilde{}}} - u_{\tilde{}} \frac{d}{dt} ln \sqrt{1-\frac{V_{qu}}{mc^2}} \qquad (62)$$

it follows that (see Appendix)

$$1+\left(1-\frac{R_g}{r}\right)^{-3} = \frac{V_{qu}}{mc^2} \qquad (63)$$

and that

$$< V_{qu} >= mc^2 < \left(1+\left(1-\frac{R_g}{r}\right)^{-3}\right) > \cong mc^2 < \left(1-\left(\frac{r}{R_g}\right)^{3}\right) > . \qquad (64)$$

Moreover, since the mass of a black hole is almost concentrated in $r << R_g$ [21, 23] (where $R_g$ is the gravitational radius), it follows both that

$$< V_{qu} >_{(n)} \approx mc^2 \qquad \forall n > 0 \, , \qquad (65)$$

and

$$\mathrm{r}_{(V_{qu(k)})} = 1+\left(1-\frac{R_g}{r}\right)^{-\frac{3}{2}} \cong 1 \qquad (66)$$

that

$$< \frac{\Lambda}{c^2} > \cong \left(\Omega_m + \Omega_{dark \;\; En}\right) < \frac{\Delta_{||(k)}}{4} > \qquad (67)$$

and, hence, by using (54) that

$$E_0 \approx < \frac{\Delta_{||(k)}}{4} > \Omega_{dark \;\; En} \;\; c^2 \frac{V_{mean}}{\mathrm{u} V_{bBH}} \approx < \frac{\Delta_{||(k)}}{4} > 8 \;\, 10^{98} J / m^3 \qquad (68).$$

Moreover, if we consider the QCD zero-point energy density $E_0$ of the vacuum [24]

$$E_0 \cong \frac{\hbar}{8f^{\,2} c^3} \check{S}^{4}{}_{max} \qquad (69)$$

from (68) it follows that

$$\hbar \check{S}_{max} \approx \left( \sqrt[4]{< \frac{\Delta_{||(k)}}{4} >} \right) 3\ 10^{18} GeV \tag{70}$$

whose order of magnitude can possibly agree with the maximum allowed Plancknian cut-off frequency

$$\hbar \check{S}_{max} = \frac{\hbar c}{l_p} = m_p c^2 = c^2 \sqrt{\frac{\hbar c}{16 f G}} \approx 5\ 10^{18} GeV\ . \tag{71}$$

where $< \frac{\Delta_{||(k)}}{4} >$ can be calculated by introducing in (7) the black hole wavefunction [21].

where $l_p$ is the Planck length.

## 5. Discussion

The results (41) basically show that in Minkowskian space-time (i.e., very far from particles) the CPTD expectation value is null regardless its zero-point energy density.

The non-zero contribution to the CC appears at second order in the GE [13]. The CPTD macroscopic expectation value is vanishing small in the region of space-time with Newtonian gravity and it increases at higher gravity as quantum-mechanical corrections. This fact well agrees [1] with the very small value of the observed CC and leads to a scenario where the major contribution to the CC comes from black holes (where the matter is so squeezed that the quantum potential energy becomes comparable with its mass energy (i.e., $V_{qu} \approx mc^2$ , and $r_{(V_{qu(k)})} \approx max\{ r_{(V_{qu(k)})} \} = 1$ ) whose gravitational effects should be primarily detectable in the motion of stars around the SMBHs at the center of the galaxies (i.e., in the bulge), in the motion of galaxies owing a SMBH an (for instance by interferometric detection of the gravitational waves [25]) in the rotation of twin black holes or neutron stars.

## 6. Conclusion

The cosmological pressure tensor density $\Lambda_{(k)Q}$ is not null if, and only if, the space-time is curvilinear and it tends to zero in the very far flat vacuum regardless its zero-point energy..

The depletion of the CPTD in the vacuum, far from material bodies, lowers its theoretical mean value on the cosmological scale, agreeing with the astronomical observations on the motion of the galaxies, by assuming the minimum allowed wave length cut-off, for the QCD zero-point energy density of the vacuum, of order of the Planck's length.

.

**Funding:** This research received no external funding.

**Conflicts of Interest:** The authors declare no conflict of interest.

# Appendix

*The hydrodynamic equation for the eigenstates of the radial symmetric Schwarzschild scalar uncharged black hole*

By using the equilibrium condition (60) and the initial condition (61) (i.e., $\chi_{(t=0)} = 1$), for the stationary eigenstates owing the radial symmetry (at least the fundamental one), equation (62) reads

$$u^{r} u^{s} \left( g_{rs} \frac{\partial}{\partial q^{\sim}} ln \left( \sqrt{1-\frac{V_{qu}}{mc^{2}}} \right) + \frac{3}{2} \frac{\partial g_{rs}}{\partial q^{\sim}} \right) = u_{\sim} \frac{d}{dt} ln \left( \sqrt{1-\frac{V_{qu}}{mc^{2}}} \right) \qquad (A.1)$$

that by using the Schwarzschild BH metric (57-59) leadsto

$$\left( g_{00} \frac{\partial}{\partial q^{1}} ln \left( \sqrt{1-\frac{V_{qu}}{mc^{2}}} \right) + \frac{3}{2} \frac{\partial g_{00}}{\partial q^{1}} \right) = u_{1} \frac{d}{dt} ln \left( \sqrt{1-\frac{V_{qu}}{mc^{2}}} \right) = 0, \qquad (A.2)$$

to

$$\frac{\partial}{\partial q^{1}} ln \left( \sqrt{1-\frac{V_{qu}}{mc^{2}}} \right) = -\frac{3}{2} \frac{\partial \, ln \, g_{00}}{\partial q^{1}} \qquad (A.3)$$

to

$$ln\left(\sqrt{1-\frac{V_{qu}}{mc^2}}\right)=-\frac{3}{2}ln\, g_{00}\,. \qquad \text{(A.4)}$$

and to

$$\left(1-\frac{V_{qu}}{mc^2}\right)=g_{00}{}^{-3} \qquad \text{(A.5)}$$